\documentclass[preprint,showpacs,preprintnumbers,amsmath,amssymb]{revtex4}

\usepackage{graphicx}
\usepackage{dcolumn}
\usepackage{bm}

\begin{document}
\title{Analysis of a Lennard-Jones fcc structure melting to the corresponding frozen liquid: differences between the bulk and the surface }

\author{   N.Olivi-Tran$^{1,2}$ and A.Faivre$^2$}
\affiliation{ $^1$S.P.C.T.S., UMR-CNRS 6638, E.N.S.C.I., 47 avenue Albert Thomas,
87065 Limoges cedex, France \\$^2$ G.E.S., Universite Montpellier II, UMR CNRS 5650
Case courrier 074, place Eugene Bataillon, 34095 Montpellier cedex 05, France}

\date{\today}

\begin{abstract}
We computed a Lennard Jones frozen liquid with a free surface using classical molecular dynamics. The structure factor curves on the free surface of this sample were calculated for different depths knowing
that we have periodic boundary conditions on the other parts
of the sample. The resulting structure factor curves show  an horizontal
 shift of their first peak depending on how deep in the sample
the  curves are computed. We analyze our resulting curves in the
light of spatial correlation functions during melting .
The conclusion is that the differences between bulk and surface are quite small during melting
and that at the end of melting, only the very surface happens to be less dense than the bulk. This result is intrinsic to the shape of  the Lennard Jones  potential and does not depend on any other parameter. 
\end{abstract}
\maketitle

\pagebreak
\section{Introduction}
The aim of this study is  to analyze the structure
of the surface of a Lennard Jones frozen liquid (within a depth of a few tens of Lennard Jones units). Our Lennard Jones
liquid may be compared to a model of glass for which the glass transition would have been obtained instantaneously from the liquid.

 Different studies deal with the analysis of glass surface.They mainly concern
the chemical nature of the surface. In the case of silicate and silica glasses, there are different numerical models, see for that the following references
 \cite{gladden,gladden2,roder,feuston,tilocca}; these previous references studied the structure, attack of water and the presence of free oxygen atoms on
 the surface of silicate glasses (with dangling bonds, as there are lacking hydrogens) as well as the variation of composition at the surface. 

But very few numerical study of the surface of simple models of liquids or glasses
have been made up to now \cite{stallons}. Stallons and Iglesia \cite{stallons}
 studied the free surface of a simple model of silica glass using the 
Stillinger Weber potential and found that the surface of their glasses was less dense than the bulk. 

We deal here with the surface
of a Lennard Jones frozen liquid. Here the action of the atmosphere and the presence or not
of dangling bonds have not been studied. We are interested in the change
in the structure of the free surface of a simple model of frozen liquid (Lennard Jones liquid) without any external action, only the fact that there are no periodic boundary conditions in one direction: the free surface.
After obtaining the Lennard Jones structure with a free surface by classical molecular
dynamics computation, we analyzed the free surface structure with the calculations
of structure factors corresponding to different depths within the sample.
This analysis is completed by the computation of the density 
as a function of depth within the sample and by computation during melting,
of the spatial correlation function in the bulk and at the surface.

The computational method used for
the sample preparation as for the calculation of the structure factor curves are presented in section II.
Section III shows the results which are discussed.
And finally, section IV is the conclusion.
\section{Numerical Procedure}
\subsection{Preparation of the numerical sample}
We used the classical molecular dynamics  to compute a Lennard Jones
liquid. The potential was a Lennard Jones 6:12 potential cut at a distance 2.5 Lennard Jones units. We began the simulation
with a fcc structure which was melt in 2.5$10^{-13}s$ (250 Molecular Dynamics-MD- steps). This is the time needed to obtain total melting
(i.e. no further melting).
The computation was done out of equilibrium: i.e. due to the free surface
we could not impose a NVT or NPT equilibrium during melting.
We chose an instantaneous cooling (in 1 MD step) in order to prevent the sample from
shearing phenomena which would rearrange the structure even at its surface.
Thus our sample is just a frozen Lennard Jones liquid.
The sample has periodic boundary conditions in all directions
except on two free surfaces on the bottom and on the top of it.
As the aim of this study is the analysis of the free surface of Lennard
Jones frozen liquids, we had to add these two free surfaces surrounded by
vacuum.
The initial box (not taking into account the boundary conditions)
has a dimension of $50 X 50 X 50$ Lennard Jones (LJ) units corresponding
to 320000 LJ particles.

In order to analyze only the action of the free surface on the corresponding
structure factors, we had to cut the initial box containing the LJ frozen liquid 
into two, according to its depth. Unless, there would be action
of the two free surfaces on the bottom and on the top of the simulated
LJ frozen liquid, and the resulting structure factor curves would be to difficult to analyze.
Furthermore it would not be a physical model of the surface of frozen liquids.
The fact that we have a free surface before having melt and frozen the sample
allows us to obtain a free relaxed surface on the top of the bulk of the 
Lennard Jones frozen liquid.
\subsection{Calculation of the structure factor}
We used a similar calculation as in reference \cite{olivitran} to obtain the structure factor
The difference is that the domain of computation is limited
by the upper surface surrounded by vacuum and a given depth within the numerical sample.
Indeed,  the wave vector may be written:
\begin{equation}
{\bf q}={ \bf k_f} - {\bf  k_i}
\end{equation}
The calculation of the structure factor curves is the following:
\begin{equation}
S({\bf q})= \frac{\epsilon ^6}{V} |\sum_{{\bf r}} [\delta ({\bf r})-c]\exp i {\bf q}. {\bf r}| ^2
\end{equation}
where $\epsilon ^6$ comes from the replacement of the integral by discrete sum
over the LJ particles, $V$ is the scattering volume (in order to normalize
all different results depending on the analyzed depth). $c$ is the concentration
of the sample which we have calculated as function of the depth
of the sample. This last feature ensures that $S({\bf q})$ tends to zero
when ${\bf q}$ tends to zero.
Then we introduce a double sum to expand the square, and, assuming isotropy
in the plane (but not in the depth of the sample), we average over all directions 
of ${\bf q}$ to get:
\begin{equation}
S(q)=\frac{\epsilon ^3}{N_{\epsilon}}\sum_{{\bf r_1}}\sum_{{\bf r_2}} (\delta({\bf r_1})\delta({\bf r_2})-c[\delta({\bf r_1})+\delta({\bf r_2})]+c^2) X \frac{\sin qr}{qr}
\end{equation}
where
\begin{equation}
r=|{\bf r_1}-{\bf r_2}|
\end{equation}
Then, transforming the double sum into a sum over all possible values for the distance $r$ and separating the $r=0$ contribution from the others, we obtain:
\begin{equation}
S(q)=\epsilon ^3(1-c+\sum_{r \neq 0} [F_a(r)-cF_b(r)]) \frac{\sin qr}{qr}
\end{equation}
where $F_a(r)$ is the mean number of LJ particles at a distance $r$ of a given
LJ particle (pair correlation function) and where $F_b(r)$ is the same quantity
for a density equal to 1.

\section{Results }
Figure 1 exhibits the typical structure factor for different depths
in LJ units. At this scale it is not possible to make any difference
between the curves. Anyway, let us remark that the curves are typical
of a glassy material with several peaks. The first peak located
at approximately at $q=0.34$ i.e. at a distance equal approximately to 3 LJ units.
This result shows that there is, in our Lennard Jones frozen liquid, a short range order up to 3 LJ units. However, the intensities of the structure
factors have been normalized by the analyzed volume (see section II.B for
details concerning this normalization). This last feature explains that we have no
vertical shift for different structure factors.

Figure 2 presents a zoom of the previous curves (figure 1) for $0.3<q<0.39$ LJ units$^{-1}$.
One may see that there is a difference for the curve corresponding
to a depth of 5 LJ units compared to the others.

Figure 3 is also a zoom of the curves presented in figure 1 but for a different
location than figure 2, here $0.24<q<0.32$ LJ units$^{-1}$. One may see again that the curve corresponding
to the depth equal to 5 LJ units is different from the others.

Finally, and in order to be able to understand our computed structure factor curves we plotted the number of LJ particles as a function of the depth from the free surface of the sample (see figure 4).
The numerical simulation box begins at height $z=1$ LJ units but as explained in section II, we cut the sample into two, so the bottom of the sample corresponds
to $z=25$ LJ units and the free surface of the sample to $z=50.1$ LJ units.

Figure 5 exhibits the distribution  of the LJ particles
 whose center of mass is contained in the slice for $49.5<z<50.1$. This may be compared to figure 6 which is the distribution
of the centers of mass of the LJ particles for $20.5<z<21.1$, i.e. for the same
width following $z$. The fact that there are zones without particles in figure 6
is explained by the width of the analyzed slice: it is less than 1 LJ unit.
In figure 5, the void zones are much larger, as expected from the variation of the number of particles presented in figure 4. What can be observed is
that distribution is rather homogeneous.
In order to go further in this analysis, we computed the two points correlation function:
\begin{equation}
g(r)=\frac{N(r)}{n^2}
\end{equation}
where $n$ is the number of particles in the slice
and $N(r)$ is the number of particles at distance $r$ from another particle.
We computed the evolution of this two points correlation function as a function of computation times of the sample structure, during melting.
 We  chose a depth of this slice equal to 0.3
because it corresponds to the depth where the surface has a lower density. Although the depth of the slice is only 0.3 LJ particles, we chose the slices at the surface
and in the bulk so that we get half of a fcc elementary cell in the slice at the beginning of computation.
Figures 7-a (left and right) are the correlation functions at the beginning of computation
(for the fcc structure) at the very surface and in the bulk. One may see a peaked distribution function. This is due to the crystalline
structure of the fcc initial distribution.
Fig.7-b,7-c and 7d correspond respectively to 100 MD time steps, 200 MD time steps and 250 MD time steps (corresponding to the frozen liquid). 
We can observe that the right and left figures 7 are very similar and we will discuss the small differences between bulk and surface later in the discussion.The correlation function
(for times t=100 MD and 200 MD  time steps ) becomes similar to
a random packing except that there are one peak
 at a distance equal to 1 LJ units.
This means that there is short range correlation during melting and that aggregates most likely remain during melting.
This peak almost disappears when the liquid is frozen for 250 MD time
steps.  Fig.7d shows that there is no more a short range correlation
and that the LJ particles are more equally distributed over the slice.

And finally, figure 8 is the plot of the number of particles per LJ unit 
volume for a slice of depth 0.3 LJ units  at the very surface and in the bulk.

\section{Discussion}
We observed that the density of our LJ frozen liquid near the surface is lower only at the very surface for a depth
included between $49.8<z<50.1$ LJ units. There is an increasing density
for $z<49.5$ LJ units and this is much less than the correlation length of about 3 LJ units (see above). Therefore the mean density for a depth of $5$ LJ units
is lower than for larger depths for which the low density of the free surface
is negligible by averaging the density over the volume.
This explains that we have no differences between the structure factors
for lower depths, i.e. $z=10,15,20$ and $25$ LJ units and very small differences for $z=5$ LJ units.
We cut the sample in the $x$ direction and checked that the density did not
vary at $x=0$ and at $x=50$ compared to the bulk. To better understand the origin of these small differences it is interesting to compare the evolution of the density at the very surface and in the 
bulk during melting.

 The evolution of the density within the bulk and at the very surface,
represented in figure 8, shows that there is not a large difference
between the two curves. This means that the LJ particles belonging
to the surface have sufficient connections with other particles
in the bulk and within the surface to prevent them to lower
the density at the very surface and to prevent them from evaporation. One can observe
that in the bulk, the density decreases during the first MD computation steps and remains
constant for all others computation times. On the other hand, there is another lowering of density of the upper surface
during the last freezing step.

Considering now the two points correlation function, and comparing the left curves of figure 7 (at the very surface) and right curves (in the bulk), we can observe a few differences. First in figure 7b , there are regular oscillations for 100MD time steps in the figure corresponding to the bulk, indicating that correlations between particles in the bulk remains on a larger time scale than at the very surface; these last disappear for 200 MD time.
In figure 7d, one can observe that at the end of melting, there are more particles at shorter distances in the bulk than at the very surface. This is well evidenced in figures 5 and 6.

Using data of figures 5 and 6, we computed the mean distance between the particles 
eliminating the void zones (i.e. : distances larger than 1.5). We obtained almost the same mean distance for the bulk and for the
surface of about 1.1 LJ unit. The particles stop to move
during melting when they reach a mean distance between particle
of 1.1 LJ unit which is the maximum of the attractive part of the LJ potential (if we eliminate
the void zones). This means that the arrangement of particles are almost the same than in the bulk, but that there is a kind of "roughness" at the very surface due to large void zones.

We can say that the free surface of a LJ frozen liquid
has also surface defects (i.e. the surface is less dense as shown
in figure 4 and as deduced from figures 2 and 3). There are some larger "voids" at the very surface
and a few particles rising above the mean surface.
But these effects concern only the region very next to the surface (for a depth
lower than 1 LJ unit). The effect of the free surface induce changes in the structure only for the first LJ particles layer.
Moreover, the bulk conserves aggregates during melting while the surface
loses this order more rapidly (in figure 7, the peak at abscissa 1 is smaller for the surface for 100MD and 200MD than for the bulk).

As a conclusion, the shape of the LJ potential leads to a surface where the density is of course
lower because of contact with vacuum. The surface has a structure of homogeneously distributed
particles which interaction is the largest attractive possible with the LJ potential surrounded by void zones:
the surface is rough.

\section{Conclusion}
We computed the structure factors as a function of the depth near the surface 
of a Lennard Jones frozen liquid. The results are that very close to the surface
the density of the numerical frozen liquid is lower. This is compared with
the analysis of the density as a function of depth into the numerical
frozen liquid and explains the shift toward lower $q$ of the structure factor curves
for lower depths. Our Lennard Jones frozen liquid is less dense near its
surface which means that the set up of the sample itself, i.e. with
a free surface leads to a rearrangement of the LJ particles without any
external action (like the attack of water on real glasses).
The interesting conclusion of our work is that even in a Lennard Jones
frozen liquid (i.e. without any action of water: presence of atmosphere or without any presence of dangling
bonds) the density and structure modification induced by the free surface
only concern a very small depth of the sample (smaller than 1 LJ unit).
During melting in the LJ fcc then LJ liquid, the LJ particle rearrange
to become very rapidly a quasi two dimensional amorphous material on the surface. 

\pagebreak

\pagebreak

\begin{figure}
\includegraphics[width=15cm]{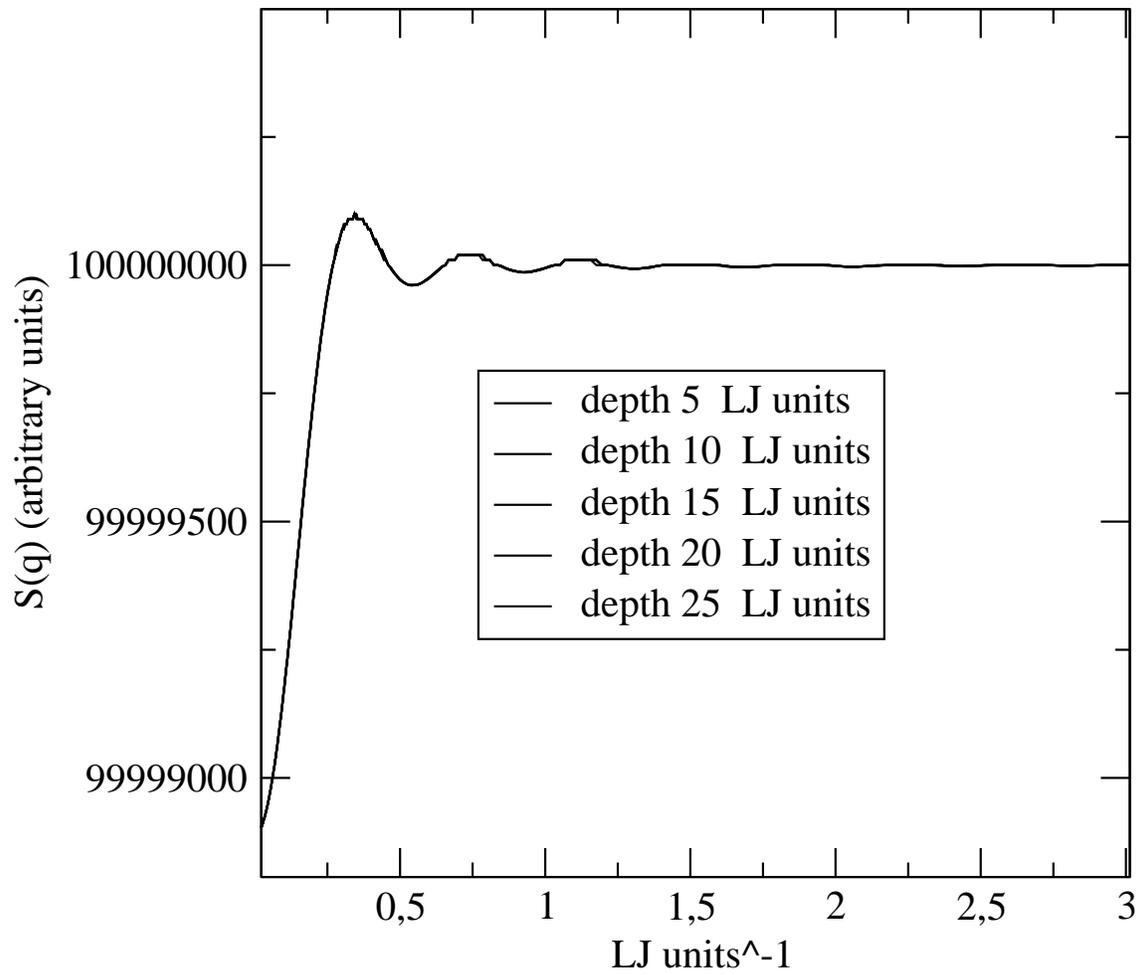}
\caption{The structure factor curves for different depths: at this scale it is not possible to make a difference between the different curves}
\end{figure}
\pagebreak
\begin{figure}
\includegraphics[width=15cm]{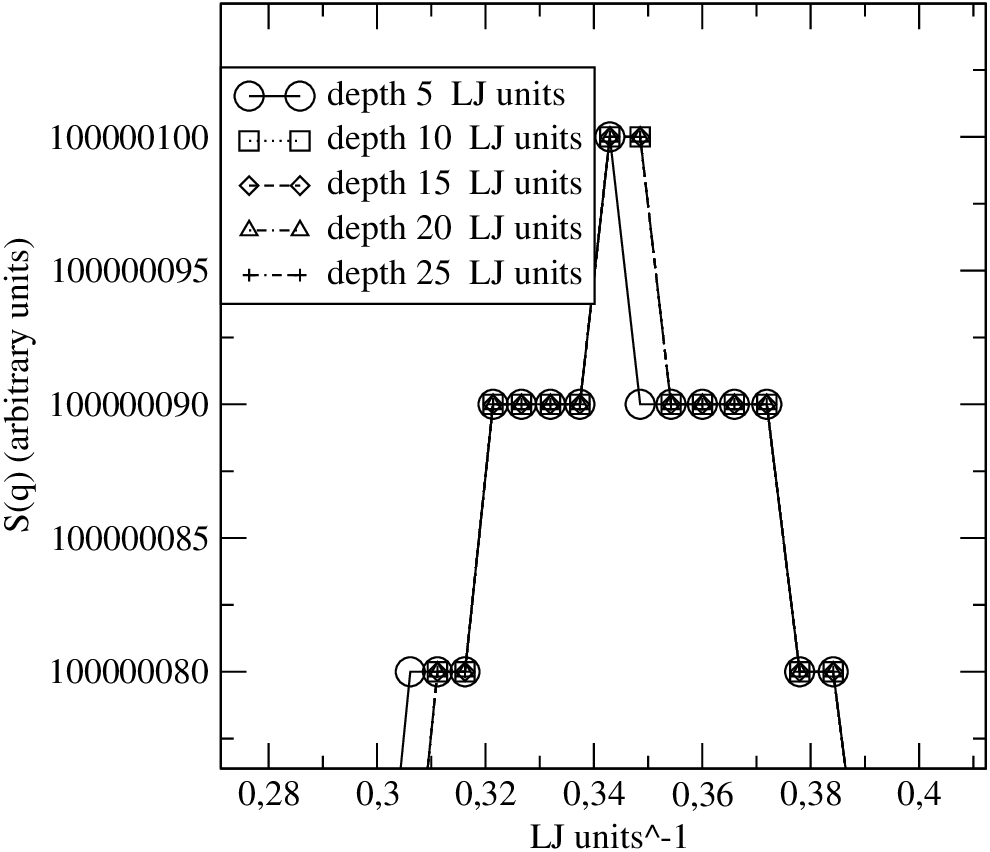}
\caption{Zoom of the different structure factor curves for different depths and for $0.3<q<0.39$ LJ unit$^{-1}$}
\end{figure}
\pagebreak
\begin{figure}
\includegraphics[width=15cm]{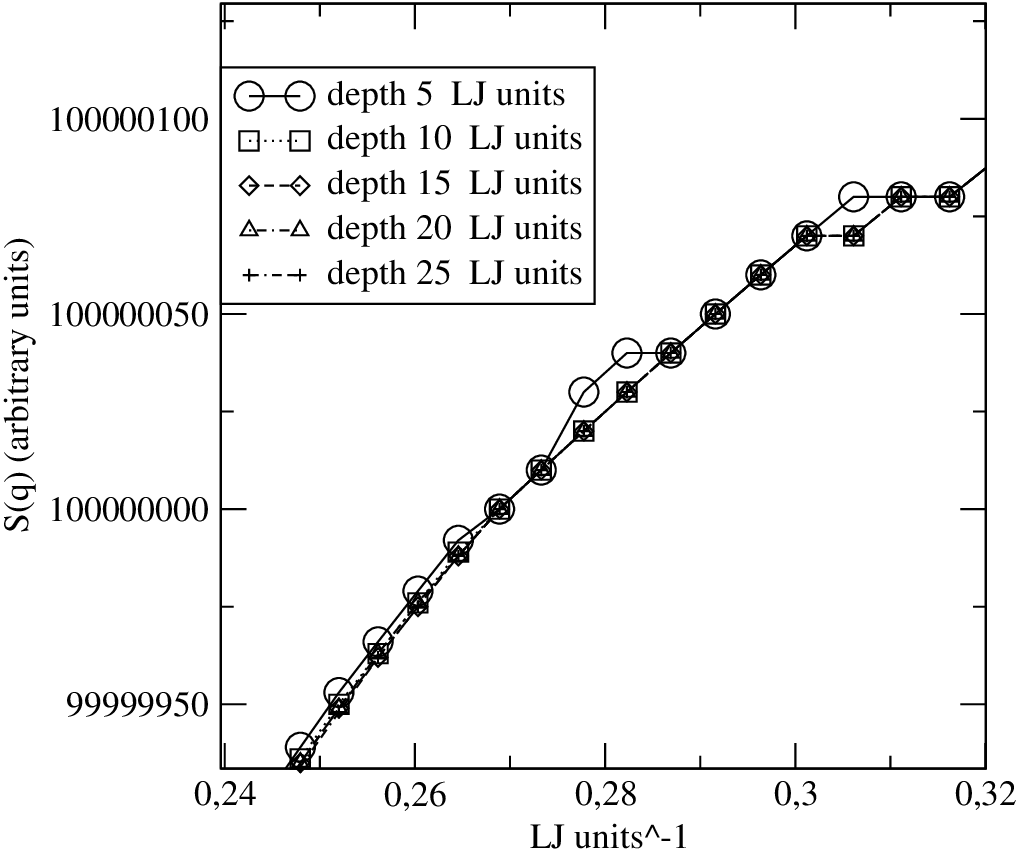}
\caption{
Zoom of the different structure factor curves for different depths and for $0.24<q<0.32$ LJ units$^{-1}$}
\end{figure}
\pagebreak
\begin{figure}
\includegraphics[width=17cm]{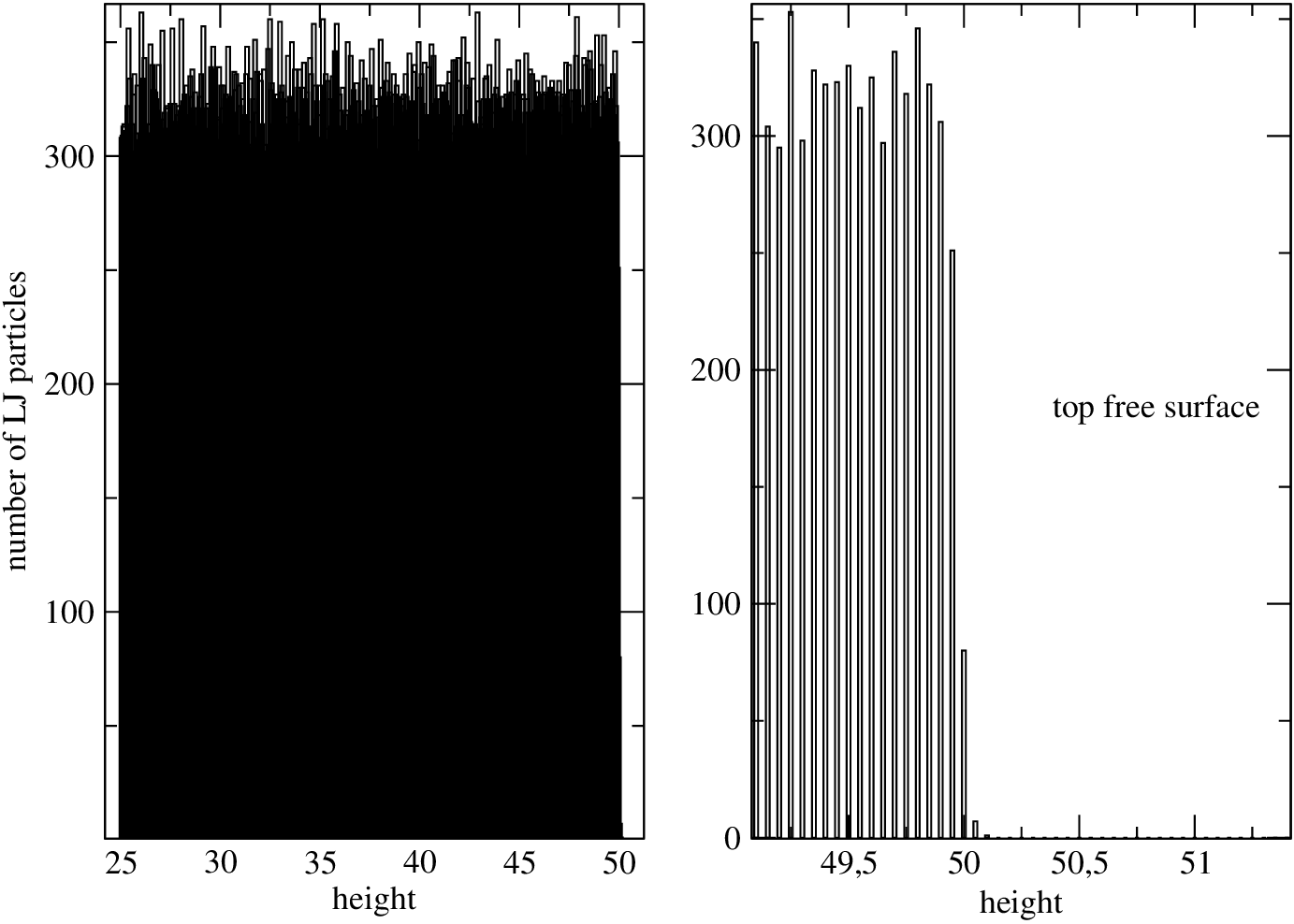}
\caption{Number of Lennard Jones particles as a function of the height of the sample.Left figure: vertical coordinates from 25 to 50.1 LJ units which is the surface of the sample, right figure: zoom on the number of Lennard Jones particles number close to the surface for a height equal to 50.1 LJ units}
\end{figure}
\begin{figure}
\includegraphics[width=17cm]{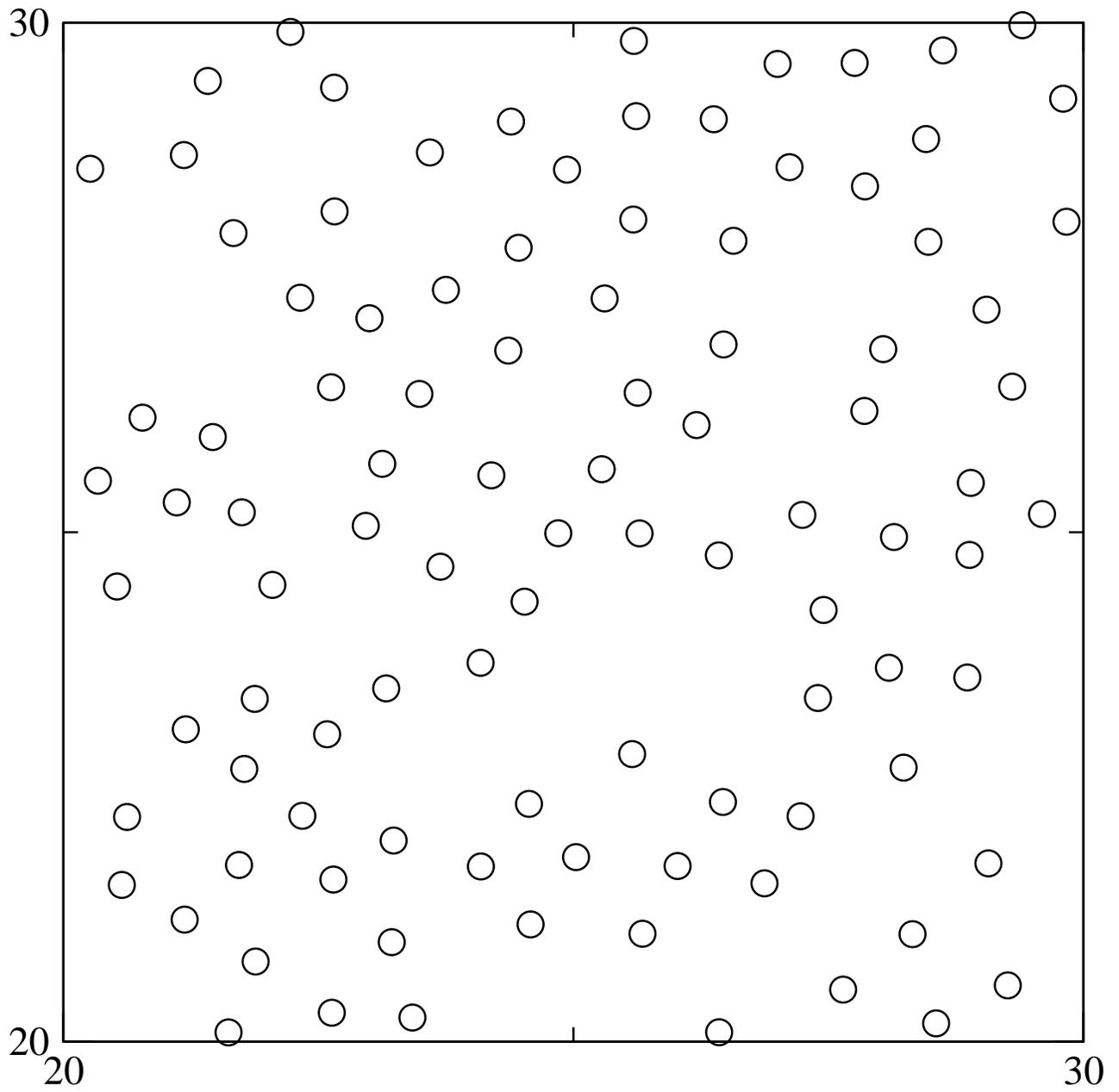}
\caption{Top view of of the centers of mass for the free surface for $49.5<z<50.1$}
\end{figure}
\begin{figure}
\includegraphics[width=17cm]{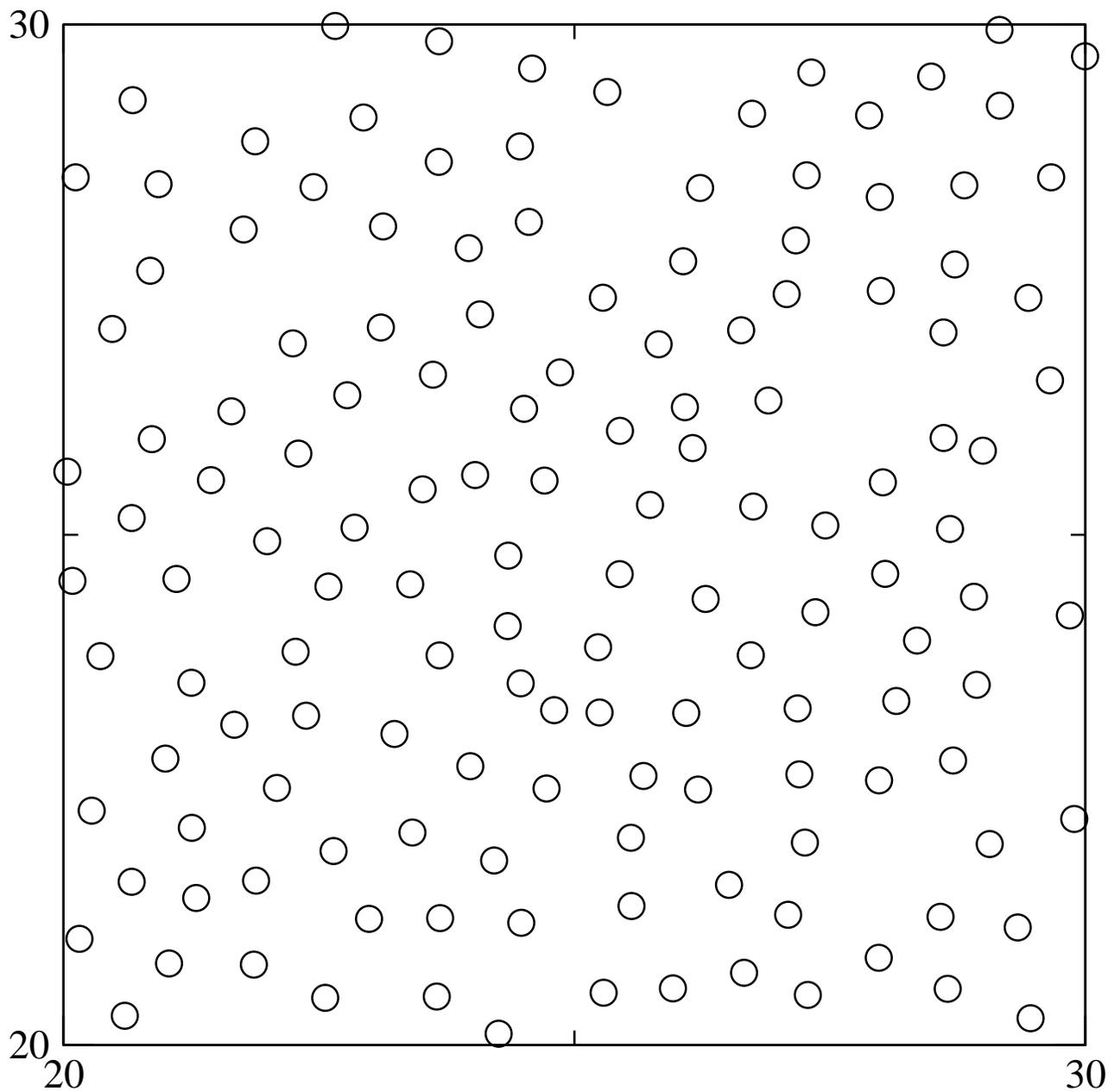}
\caption{Top view of the centers of mass of the LJ particles for the bulk for $20.5<z<21.1$}
\end{figure}
\begin{figure}
\includegraphics[width=17cm]{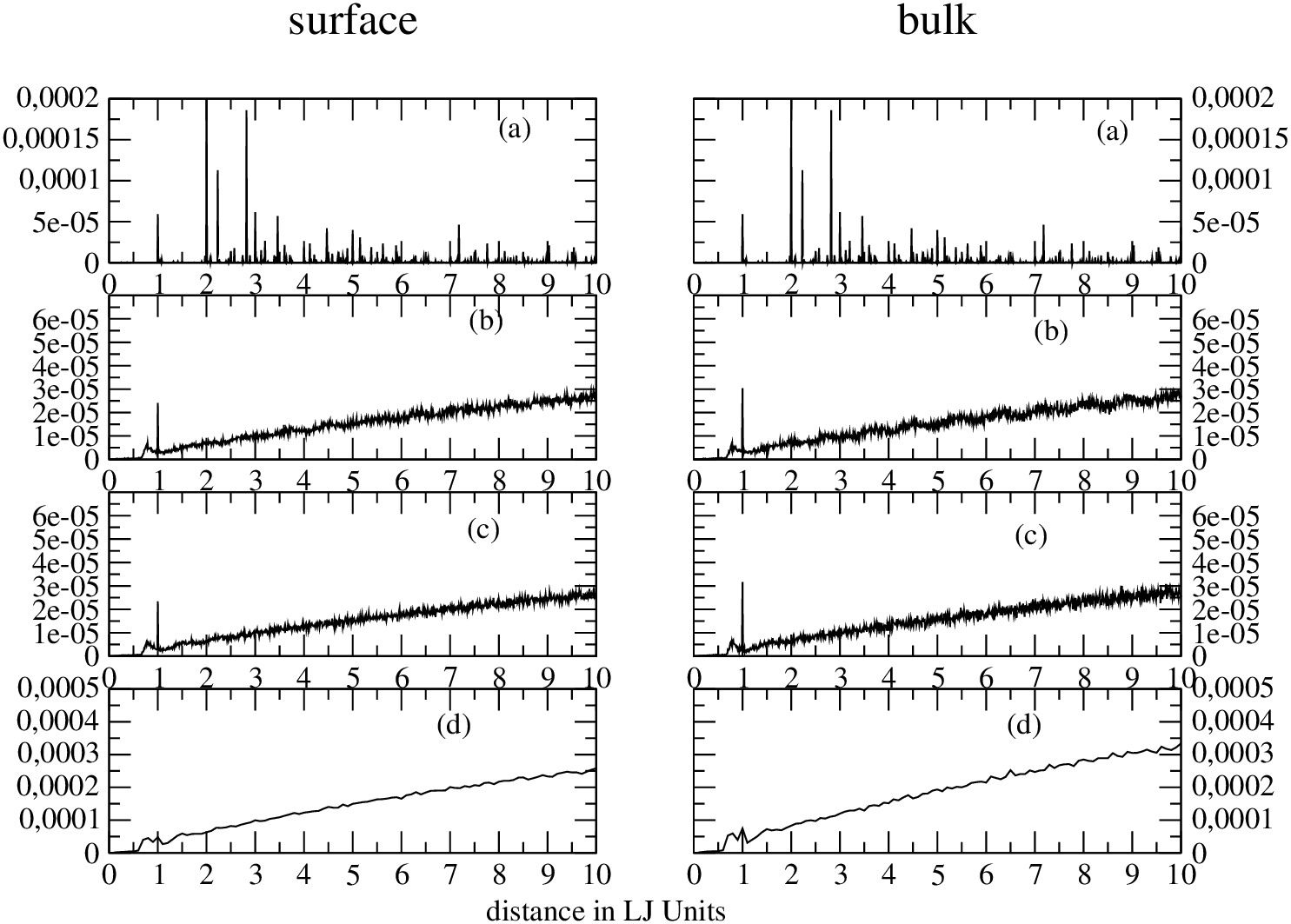}
\caption{Two points correlation function for a slice at the free surface (left) and in the bulk (right)
within a depth of 0.3 LJ units. (a): for the fcc structure at the beginning of computation (0 MD time step); (b): for 100 MD time steps; (c): for 200 MD time steps; (c) for 250 MD time steps when the liquid is frozen}
\end{figure}

\begin{figure}
\includegraphics[width=17cm]{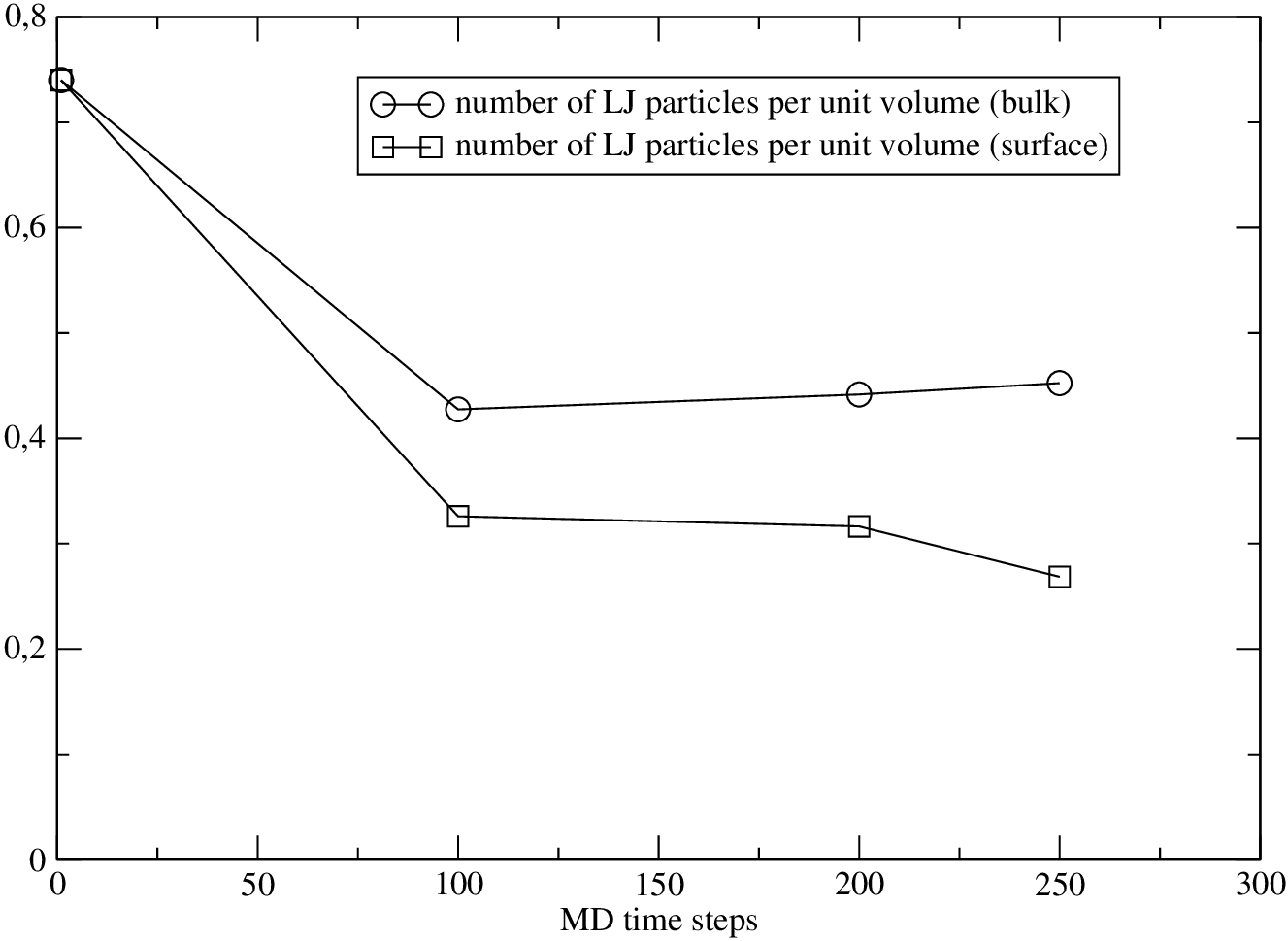}
\caption{Number of LJ particles per LJ unit volume in the bulk and at the very surface both for a depth of 0.3 LJ units}
\end{figure}
\pagebreak

\end{document}